\documentclass[referee]{aa} 
\usepackage{graphicx,psfig}
\begin{document}

\title{Pulsar Luminosity Function}
\titlerunning{Pulsar Luminosity Function}  
\author{O. H. Guseinov\inst{1,2}  
\and 
E. Yazgan\inst{2}
\and
S. \"{O}zkan\inst{1}
\and
S. Tagieva\inst{3}}
\authorrunning{Guseinov et al.}
 
\offprints{E. Yazgan}

\institute{Department of Physics, Akdeniz University, Antalya, Turkey \\
\email{huseyin@pascal.sci.akdeniz.edu.tr}
\and
Department of Physics, Middle East Technical University,
Ankara 06531, Turkey \\
\email{yazgan@astroa.physics.metu.edu.tr} 
\and
Academy of Science, Physics Institute Baku 370143, Azerbaijan Republic \\
\email{msalima@lan.ab.az}
}

\date{Received; accepted}

\abstract{We construct and investigate the pulsar luminosity function 
using the new catalogue which includes data for 1315 radio pulsars. The 
luminosity 
functions are constructed for 400 and 1400 MHz separately, and they are 
compared. Also, the luminosity functions excluding the binary millisecond 
pulsars and other pulsars with low magnetic fields are constructed. The 1400 
MHz 
luminosities as a function of characteristic age and as a function of 
magnetic field for radio pulsars, anomalous X-ray pulsars and dim radio 
quiet neutron stars are presented and the implications of the pulsar 
luminosity function on these new kind of neutron stars are discussed. 
 }

\maketitle

\section{Introduction}

As known, radio pulsars (PSRs) do not radiate isotropically, but radiate in a
beam. The beaming angle in the radio band is much less than the ones in  
the X-ray
band (Lyne \& Graham-Smith 1998). Now, we have the data for 1315 PSRs 
(Guseinov et al. 2002a).
From these PSRs, about 30 of them with characteristic ages $\tau <$10$^7$ 
yr are observed also in optical, X-ray and
$\gamma$-ray bands. Becker \&
Tr\"{u}mper (1997) gives the list for 19 radio PSRs with ages $\tau <$10$^7$
yr observed also in the X-ray band with the $ROSAT$ satellite.   

For the last 10 years, very important objects in astrophysics;  
so called soft gamma repeaters (SGRs), anomalous X-ray pulsars (AXPs) and
dim radio quiet neutron stars (DRQNSs) has been investigated very 
intensively (Mereghetti et al. 1996; Kaspi et al. 1996; Thompson \& 
Duncan 1995; Brazier \& Johnston 1999; Halpern et al. 2002). There is no 
radio radiation observed 
from these sources (Gaensler et al. 2001).  Some of these single neutron 
stars, namely, SGRs and
AXPs have X-ray spin periods between 5-10 s and their magnetic field
strengths are believed to be B=10$^{14}$-10$^{15}$ G (Thompson \& Duncan,
1995). It is believed that
the age of AXPs and SGRs absolutely are not more than 10$^5$ yr
(Tagieva \& Ankay, 2002). Since PSRs with
characteristic time $\tau < 10^5$ yr have magnetic fields mostly
of 10$^{12}$-10$^{13}$ G and since no PSR has a magnetic field as high as
10$^{14}$-10$^{15}$
G, AXPs and SGRs are named as magnetars. It is important to note that now
the properties of AXPs and SGRs are mostly investigated in the frame of
the magnetar model, however, we do not know definitely that these neutron
stars have such huge magnetic fields. We wonder whether the non-detection of
radio radiation from SGRs, AXPs and DRQNSs with spin periods P$\sim$5-10 
s is due to the strong magnetic fields?

There are about 10 dim X-ray point sources. Some of these sources with
small periods are connected to supernova remnants (SNRs) (see 
Gaensler et al. 2001 and Tagieva \&
Ankay 2002 for review). There is no radio radiation detected
from these sources in spite of their small distances from the Sun.
We may expect that the reason
that there is no radio radiation from DRQNSs with small periods is   
because of the radio luminosity function.

For most of the SNRs closer than 5 kpc, there have
been searches for pulsars (Gorham et al. 1996, Lorimer
et al. 1998). However, it has been very hard to detect them.
Up to 5 kpc from the Sun, only 6-7 pulsars are found to
be genetically connected to SNRs (Kaspi \&
Helfand 2002, Camilo et al. 2002, Guseinov et al. 2002b). Most of them are
discovered in surveys. There are about 79 SNRs up to the distance 5 kpc 
(Guseinov et al. 2002b).   This shows that on the average, for SNRs closer 
than 5 kpc, only 1
genetic connection exists for 11-13 SNRs. Is not this a result of the
luminosity function and background radiation? In this paper, we aim at
finding the improved luminosity
function at 400 MHz and for the first time at 1400 MHz which is more 
important today due to the fact that the radio observations of AXPs, 
SGRs and DRQNSs are conducted at 1400 MHz. We also aim at investigating the 
properties 
of the luminosity function for all PSRs and single born PSRs, separately.

\section{Luminosity Function For All Pulsars at 400 and 1400 MHz}
In the last 30 years, PSR luminosity functions are constructed for many
times.
Most of them have been done by different authors between the years
1975-1981. From these works which are appropriate, it has been found that
the 400 MHz luminosity function has a power law form with slope -(1.6-2.1)
(see Guseinov et al. 1982 for a review). After the publication of  
Taylor et al. 
(1996) catalogue, Lorimer et al. (1993) constructed the luminosity function
for L$_{400} > 10$ mJy kpc$^2$ and Allakhverdiev et al. (1997)
constructed the luminosity function for L$_{400} > 1$ mJy kpc$^2$.
Allakhverdiev et al. (1997)
found the slope of the luminosity function to be -0.9 which is consistent
within the error
limit of the slope given in Lorimer et al. (1993). Evidently, the error in
the luminosity
function is high due to the fact that the number of PSRs with the highest and
the lowest luminosities is few.

From 1315 PSRs, for 685 of them the flux value exists at
400 MHz, and for 862 of them the flux value exist at 1400 MHz (Guseinov et
al. 2002a). The number of   
PSRs having flux data measured at 1400 MHz sharply increased because of  
the observations of the sky
(mostly the galactic plane in the southern sky) is conducted at that 
frequency in the last years (Lyne et al. 2000; Camilo et al. 2000; 
D'Amico et al. 2001, Edwards \& Bailes 2001).
The number of PSRs with Log L$_{1400}$ and Log L$_{400}$ less than zero
have considerably increased. 
We have indicated the importance of the luminosity function for PSRs with
low luminosity in the introduction. Figure 1 displays the Log L$_{400}$ vs.
distance distribution for PSRs closer than 1 kpc. In the volume up to 1
kpc, L$_{400}$ values are known for 70 PSRs. For 12 of them Log L$_{400}
< 0$.

In Figure 2, Log L$_{1400}$ vs. distance is displayed for PSRs closer than
1.5 kpc. In the volume up to 1.5 kpc, L$_{1400}$ is measured for 101 PSRs and
33 of them have Log L$_{1400} < 0$.
No farther PSRs are known with such low luminosity values in the Galaxy.
For both of the frequencies there are enough number of PSRs with low
luminosity. This provides us with the opportunity to construct the
luminosity
functions for both frequencies down to low luminosities realistically.  
Figure 3 displays the luminosity function for PSRs with known flux
values at 400 MHz. In Figure 3, N represents the
number density of PSRs which have luminosity values equal to or greater than
the luminosity value corresponding to N.
As seen from the figure, there is not one linear relation between 
Log N and Log L$_{1400}$.
We have fitted the luminosity function by three different linear
approximations. Log N dependence on Log L$_{400}$ has a sharp slope for
Log L$_{400} >$ 1.5 and have the expression:
\begin{equation}
N=520 L_{400}^{-0.81 \pm 0.01}
\end{equation}
The portion of the luminosity function for luminosity values in the range 
0.2$<$Log L$_{400} <$1.5 is fitted by the equation:
\begin{equation}
N=62 L_{400}^{-0.19 \pm 0.01}
\end{equation}
and the slope for lower luminosity values (-1$<$Log L$_{400} 
<$0.2) is given by the expression:
\begin{equation}
N=57.5 L_{400}^{-0.071 \pm 0.006}
\end{equation}
The slope of the dependence (1) in error limits is consistent with slopes
given in Lorimer et al. (1993) and Allakhverdiev et al. (1997). But the
slope of the luminosity function for Log L$_{400} <$1.5
continuously gets more flat, therefore, the number of PSRs with
small luminosities do not increase as strong as we think before.

Figure 4 displays the luminosity function for PSRs with flux values
known at 1400
MHz. Again, we have fitted the luminosity function with three different 
lines for three different luminosity intervals. The high luminosity part
(Log L$_{1400} > 0.5$) of the luminosity function is fitted by the 
expression: 
\begin{equation}
N=188.4L_{1400}^{-0.95 \pm 0.01}
\end{equation}
In the interval -0.5$<$Log L$_{1400}
<$0.5, the fitting function becomes:
\begin{equation}
N=85.1L_{1400}^{-0.27 \pm 0.01}
\end{equation}
and at smaller values of luminosity (-1.5$<$ Log L$_{1400} 
<$-0.5), the luminosity function have a smaller slope given by the 
expression: 
\begin{equation}
N=101 L_{1400}^{-0.13 \pm 0.01}
\end{equation}

The data points in Figure 3 and 4 are calculated as the average of PSR
number density in three different volumes.
These volumes are chosen such that there are enough number of PSRs for a
fixed luminosity, so that the largest volume contains up to the most
luminous PSRs and the smallest volume contains the least luminous PSRs.
The errors are calculated as the
deviation of these values from the average values for these three 
volumes.

\section{Luminosity Function for Single Pulsars with Characteristic Ages
$\tau < 10^7$ Years}
As mentioned in the introduction the radio fluxes for SGRs and AXPs are
unknown or for some of these sources upper limits are known (Gaensler et
al. 2001). This is also true for DRQNSs
(Halpern et al.
2002; Mereghetti et al. 1996; Vasisht et al. 1997; Brazier \& Johnston
1999; Petre et al. 1996; Walter et al. 1996; Neuhauser et al. 1997; Kaspi
et al. 1996; Motch 2000; McLaughlin et al. 1999, 2000; Roger et al. 1988,
Kaspi et al. 1998; Hambaryan et al. 2002; Lorimer et al. 1998; Seiradakis 
1992;
Kassim \& Lazio 1999; Shitov et al. 1997; Malofeev \& Malov 1997;
Ramachandran et al. 1998).
If for SGRs and
AXPs the
non-existence of radio luminosity is due to the very high magnetic fields,
then
what is the reason that DRQNSs with small periods have very low radio  
luminosity? For most
of the single neutron stars born with magnetic fields of
10$^{11}$-10$^{13}$ G,
PSR luminosities seem not to be related with the strength of
the magnetic fields of such single PSRs. We wonder whether there are
differences in the
radio radiation properties of PSRs and SGRs/AXPs/DRQNSs and can their radio
luminosity be predicted from the luminosity function of PSRs? 
To answer these questions and to avoid the influence of the binary 
millisecond PSRs, we constructed the luminosity function for the single 
born PSRs.

To exclude the binary millisecond PSRs and other PSRs with small values 
of magnetic field strength, we took the PSRs having $\tau <$10$^7$ yr and 
constructed the luminosity function. The number of
PSRs with $\tau <$10$^7$ yr and having flux observed at 400 MHz is
364, but in the region up to d$<1.5$ kpc from the Sun, their number is
only 44.
The luminosity function for this sample of PSRs is given in Figure 5. As
seen from the
figure, for Log L$_{400} <$ 0.5, the slope of the luminosity function 
approaches to zero. For higher luminosity values (Log L$_{400} >$ 1.5), the
slope of the luminosity function is given by the expression:
\begin{equation}
N=158.5L_{400}^{-0.63 \pm 0.02}   
\end{equation}
But the portion of the luminosity function
in the interval 0.4$<$Log L$_{400} <$1.5
can be fitted with the expression:
\begin{equation}
N=52.5L_{400}^{-0.31 \pm 0.01}
\end{equation}

The total number of
PSRs with $\tau <$10$^7$ yr and having flux observed at 1400 MHz is
562 but in the volume around the Sun with radius of d$<$1.5 kpc is 35.
As seen from Figure 6, the part of the luminosity function with 
luminosity values Log L$_{1400} >$ 0.3 is fitted by:
\begin{equation}
N=66 L_{1400}^{-0.75 \pm 0.10}
\end{equation}
For lower values of luminosity Log L$_{1400} <$0.3, the luminosity 
function has a slope given by the equation: 
\begin{equation}   
N=40 L_{1400}^{-0.05 \pm 0.01}
\end{equation}
The luminosity function for PSRs with fluxes known at 1400 MHz is
constructed for
the first time, therefore there is no other luminosity function published
for comparison.

\section{Discussion and Conclusion}
In Sections 2 and 3, the luminosity function is found for PSRs with
luminosity known at 400 and 1400 MHz. Luminosity functions are found both
for all PSRs and
for PSRs which are born single and have characteristic ages $<$10$^7$ yr. 
Before, the luminosity function for 1400 MHz could not be
constructed because the number of PSRs observed at this frequency was
few. Now since the number of such PSRs are 862 in the Galaxy (Guseinov et
al. 2002a), we constructed the luminosity function at 1400 MHz. To compare
the luminosity functions in Figures 3-6, we brought them together in Figure
7. As
known, using the luminosity function, the ratio of space density of PSRs
with different
luminosities can be determined. To find the real density of PSRs in any
luminosity interval or PSRs with luminosities higher than any given
luminosity, the
function must be calibrated by considering the number density of PSRs
which have luminosity higher than a
chosen luminosity value.
In Figure 7, we calibrated the luminosity functions such that they have the
same values at Log L=-1. Using this function we may have information 
about the relative number of PSRs with different values of luminosity values.

As seen from Figure 7, the luminosity function for single born PSRs 
(lines 5
and 6) and luminosity function
for all PSRs (lines 3 and 4) are similar at the same frequencies. There
is huge difference when we compare the luminosity functions at 400 and 1400
MHz.
The reason of this huge difference is that the spectral index of PSRs is
sharp. By comparing the lines 3 and 5 in the figure, we see that
the number density of PSRs which are born single
with Log L$_{400}$ close to 1.5 is less but number density of single PSRs 
with Log
L$_{400} \sim$ 0-0.5 is more. On the other hand, if we compare the lines 4
and 6 we see that single
born PSRs with Log L$_{1400} >$0.5 is more.
There might be some doubts in the comparison of the line 3 and 5 which 
are closer to each other, however, the results obtained by comparison 
of   
the line 4 and 6 are must be trustworthy.

If the AXPs and SGRs are magnetars then the non-detection of radio 
radiation from them must be explained. If AXPs are accreting
neutron stars, then they will not emit any radio radiation. The upper
limits of radio fluxes at 1400 MHz for AXPs are known (Gaensler et al. 
2001). In Figure   
8 and 9, Log L$_{1400}$ values versus Log $\tau$ and Log B for PSRs with
ages $<$3 10$^5$ yr are represented. As seen from these figures, the
radio luminosities of PSRs practically do not depend on $\tau$ or B. Upper 
limit   
values for radio luminosities of 4 AXPs on the average are smaller than 
that   
of young PSRs. AXPs 1E2259+586 and 4U0142+625 have very small upper
limits for luminosity. As we see from Figure 6 number of PSRs with such
small luminosity must be very small. Therefore we may believe that
AXPs and may be also SGRs have very small radio luminosity.

In Figures 8 and 9 also upper limit for radio luminosities of DRQNSs are   
presented. As seen from figures, without any doubt, this class of neutron
stars practically do not have radio radiation. 
The Number of PSRs younger than 10$^5$ yr up to the distance 1 kpc is 2. 
The number of DRQNSs with such ages and distances is also 2. 
This shows that the 
birth rate of DRQNSs in the Galaxy must be closer to the birth rate of 
radio PSRs. This situation does not change if we take the beaming factor 
into account. The beaming factor for radio PSRs is 1/2 which implies the 
number of radio PSRs younger than 10$^5$ yr and closer than 1 kpc to be 4.
$CHANDRA$ satellite observations will considerably increase the number of 
known DRQNSs.

\newpage

\begin{figure*}
\centerline{\psfig{file=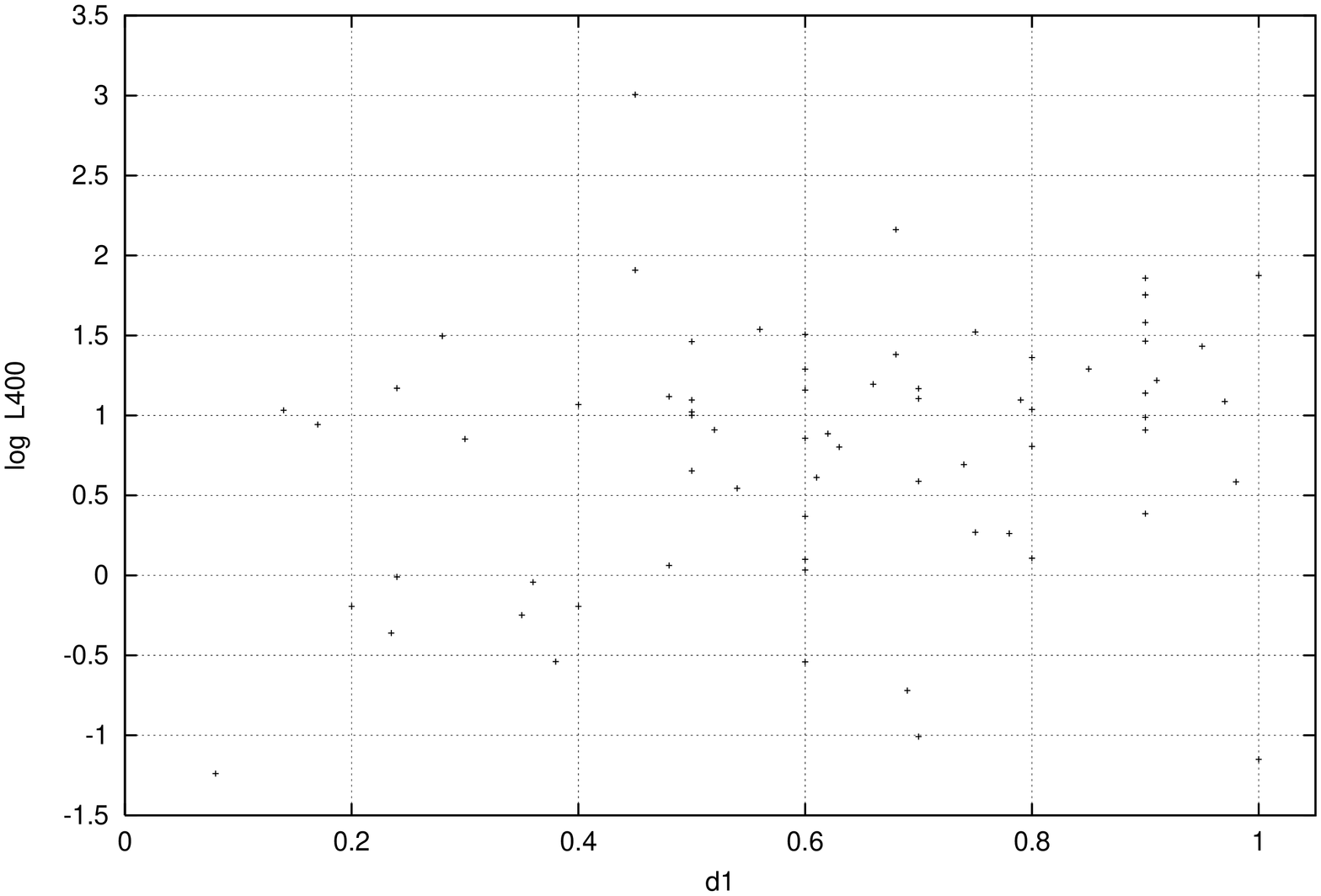,width=13cm,angle=-90}}
{Figure 1. Distribution of luminosity of PSRs at 400 MHz (L$_{400}$) vs.
distance values from the Sun for 70 PSRs up to 1 kpc.}
\end{figure*}

\begin{figure*}
\centerline{\psfig{file=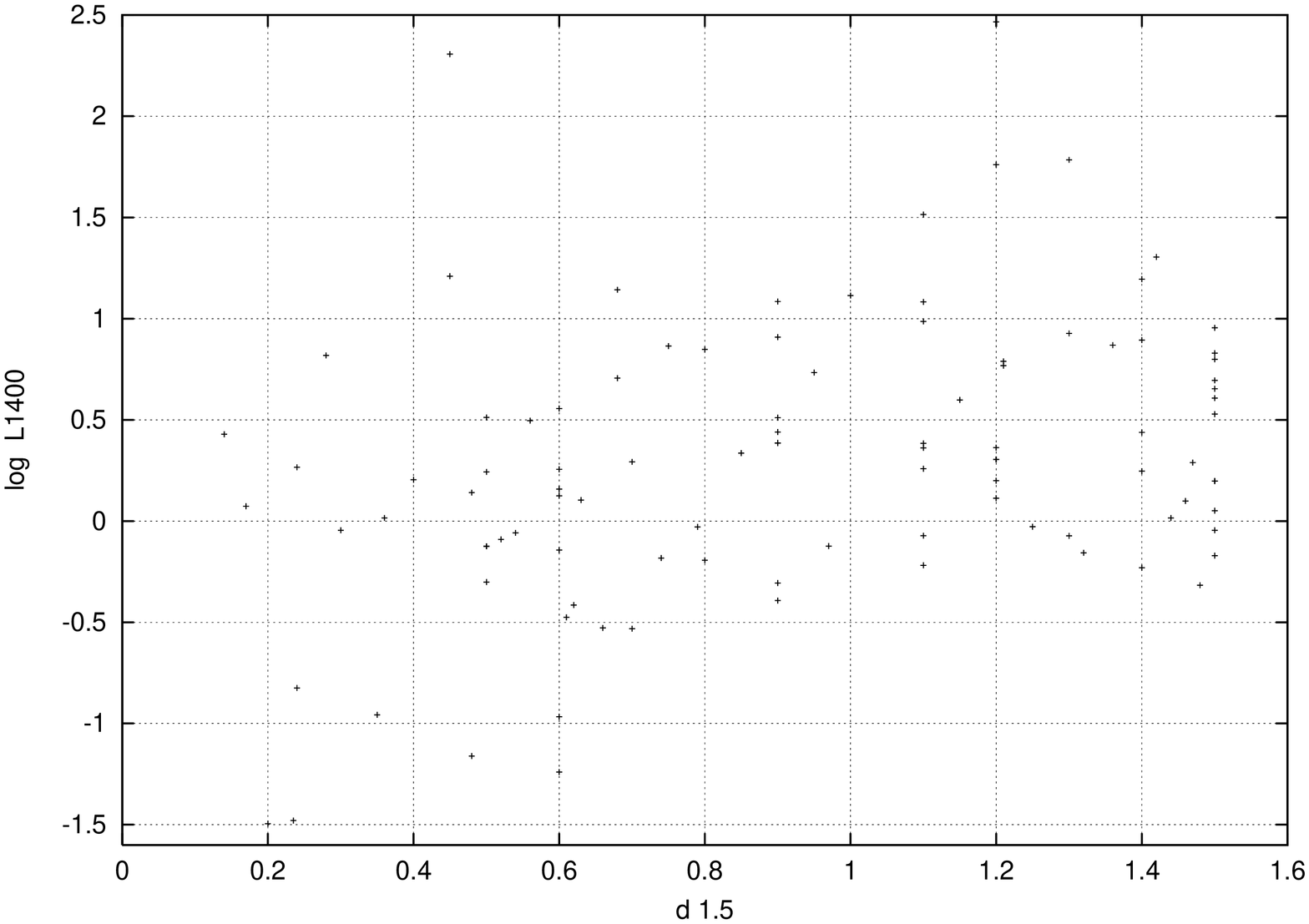,width=13cm,angle=-90}}
{Figure 2. Distribution of luminosity of PSRs at 1400 MHz (L$_{1400}$ vs.
distance from the Sun for 101 PSRs up to 1.5 kpc}
\end{figure*}  

\begin{figure*}
\centerline{\psfig{file=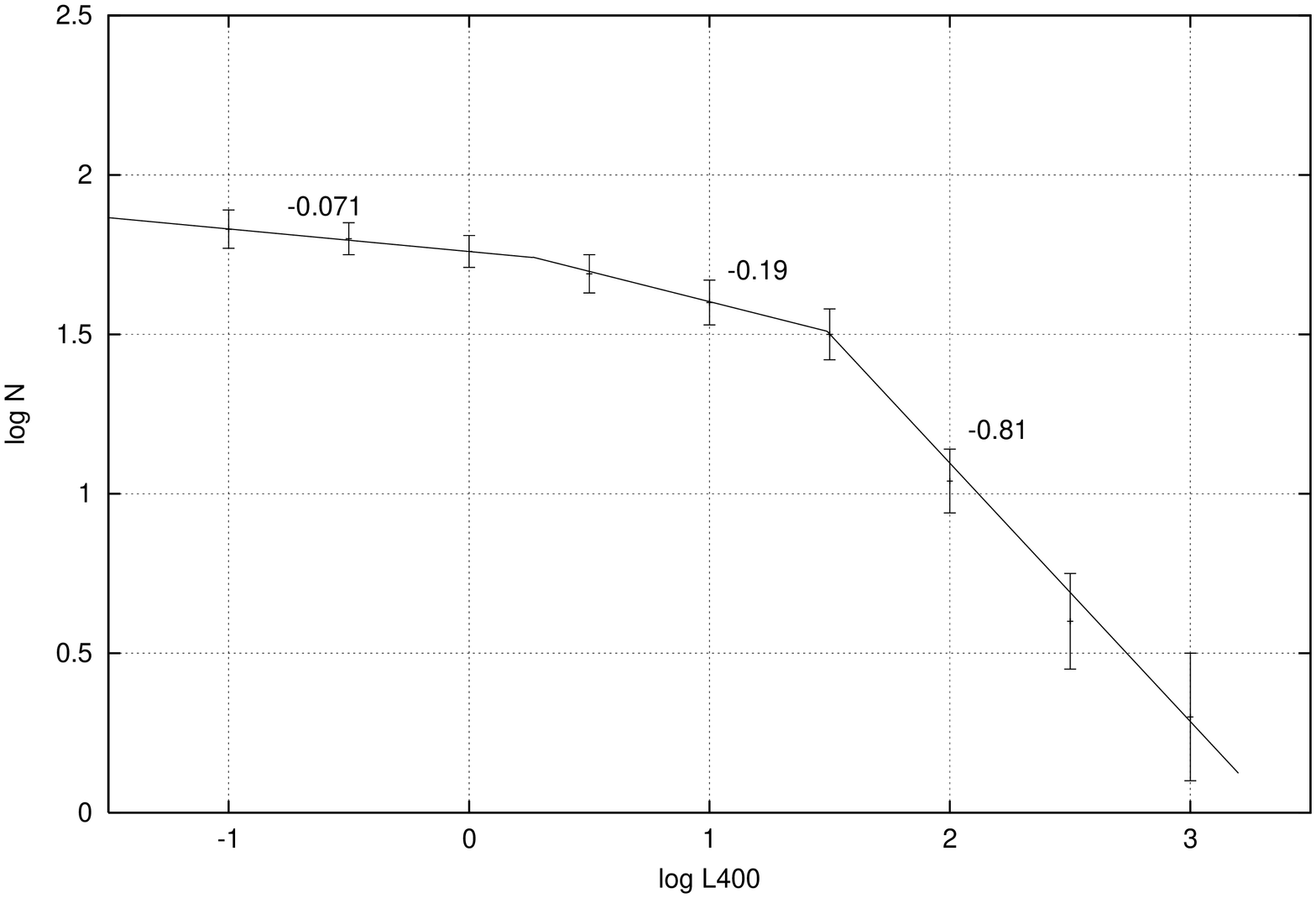,width=13cm,angle=-90}}
{Figure 3. Log N - Log L$_{400}$ dependence for all PSRs. The numbers in 
the figure represent the degree of the slope.}
\end{figure*}

\begin{figure*}
\centerline{\psfig{file=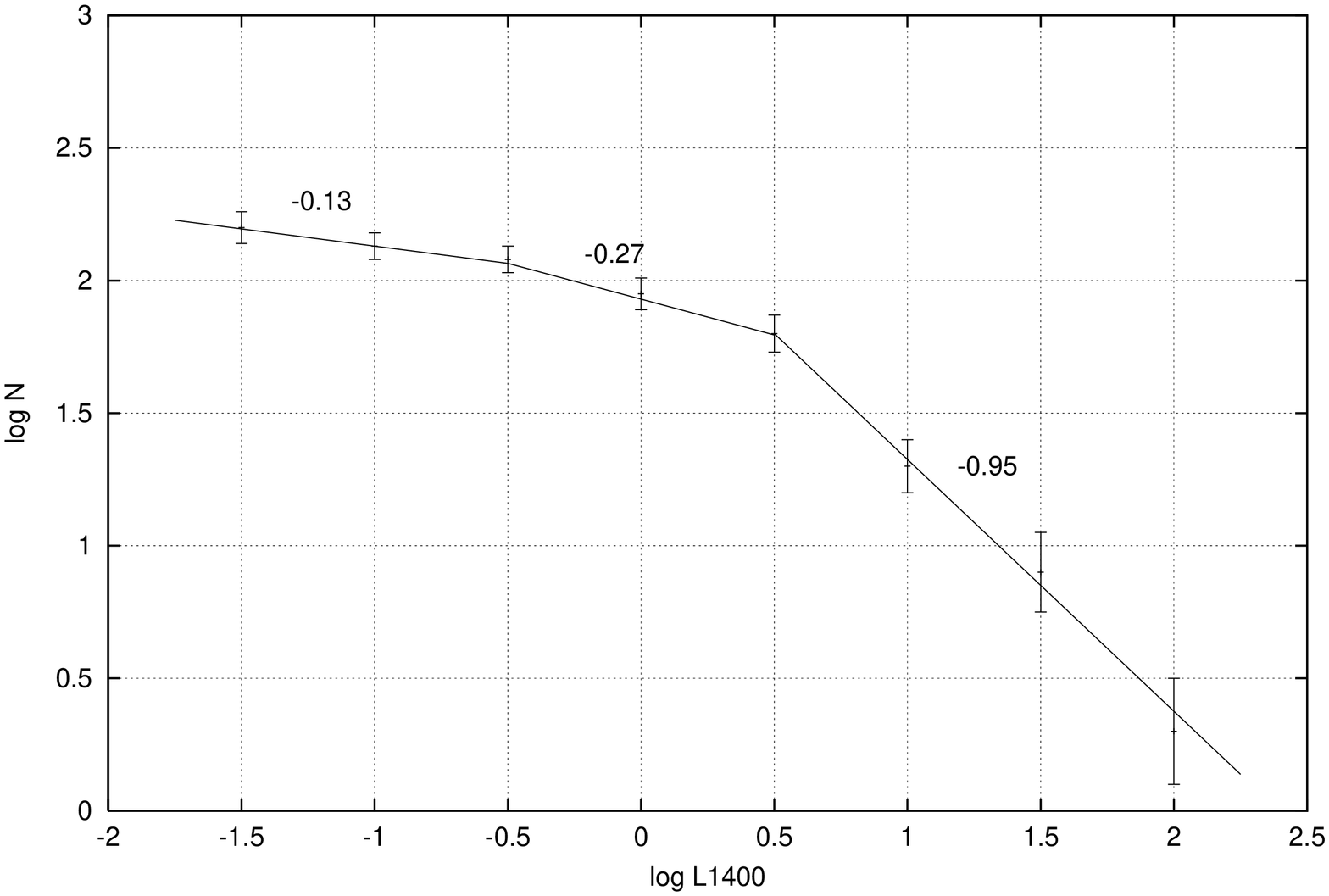,width=13cm,angle=-90}}
{Figure 4. Log N - Log L$_{1400}$ dependence for all PSRs. The numbers in
the figure represent the degree of the slope.}
\end{figure*}

\begin{figure*}
\centerline{\psfig{file=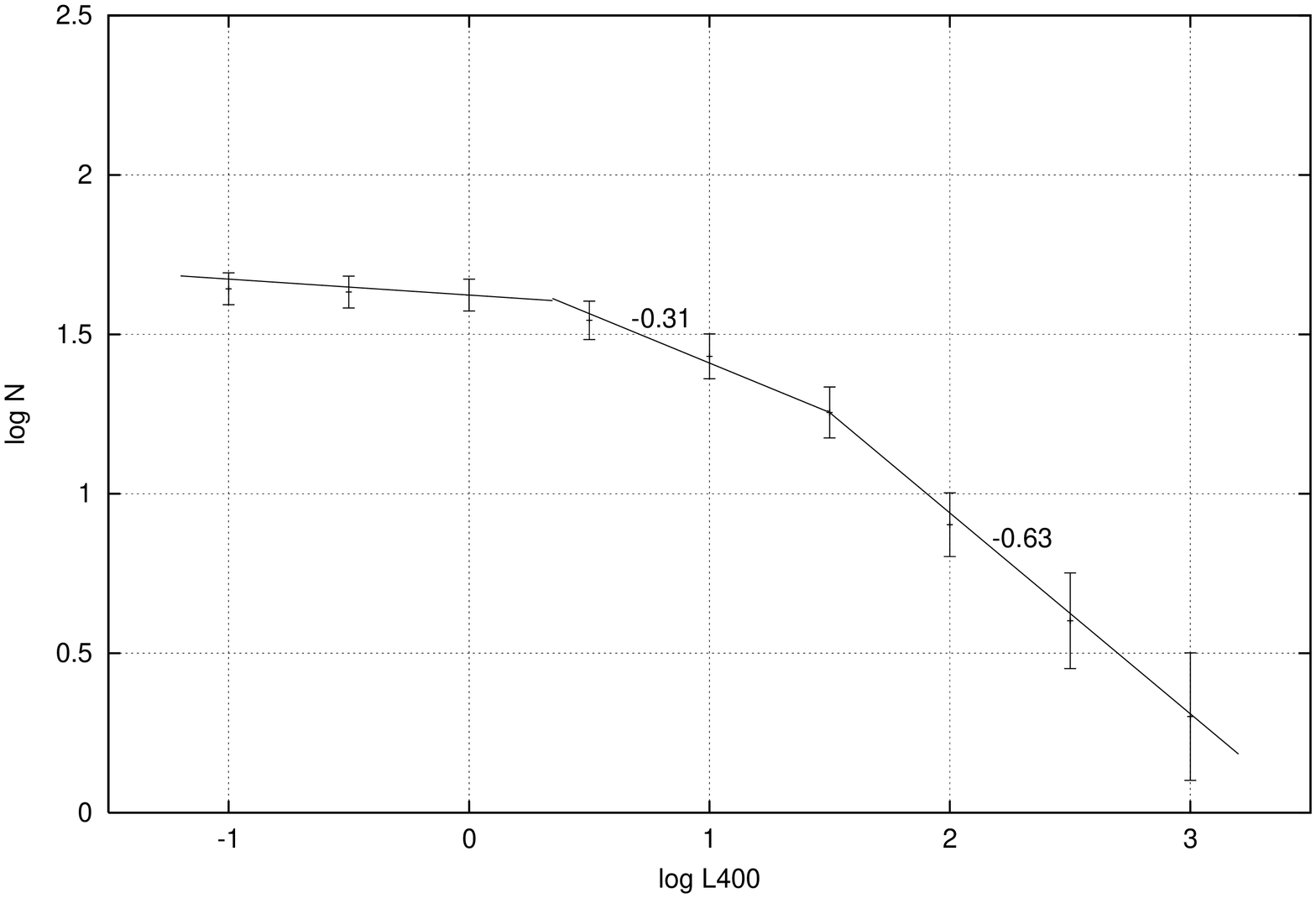,width=13cm,angle=-90}}
{Figure 5. Log N - Log L$_{400}$ dependence for single PSRs with 
characteristic ages $< 10^7$ yr. The numbers in
the figure represent the degree of the slope.} 
\end{figure*}

\begin{figure*}
\centerline{\psfig{file=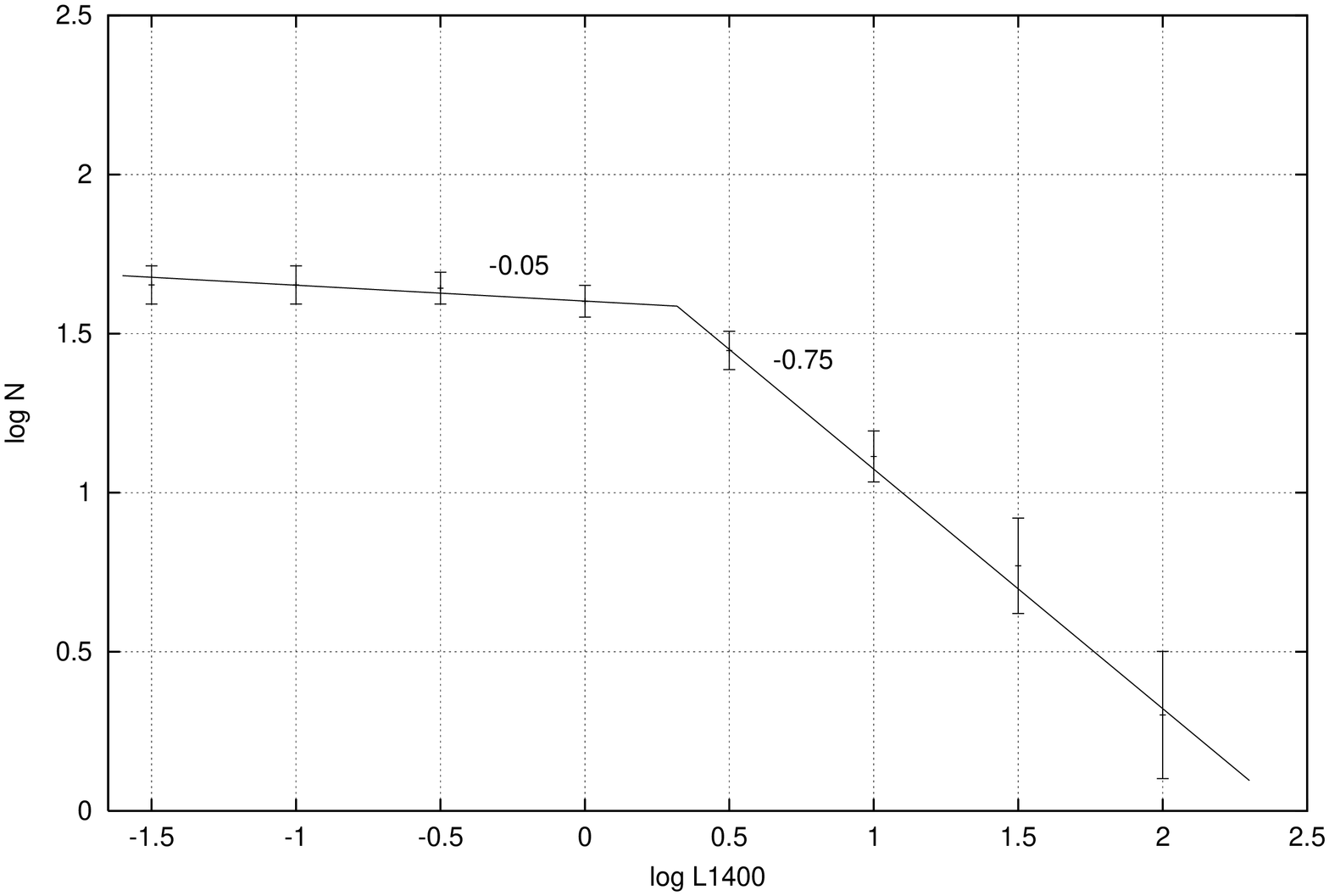,width=13cm,angle=-90}} 
{Figure 6. Log N - Log L$_{1400}$ dependence for single PSRs with 
characteristic ages $< 10^7$ yr. The numbers in
the figure represent the degree of the slope} 
\end{figure*}

\begin{figure*}
\centerline{\psfig{file=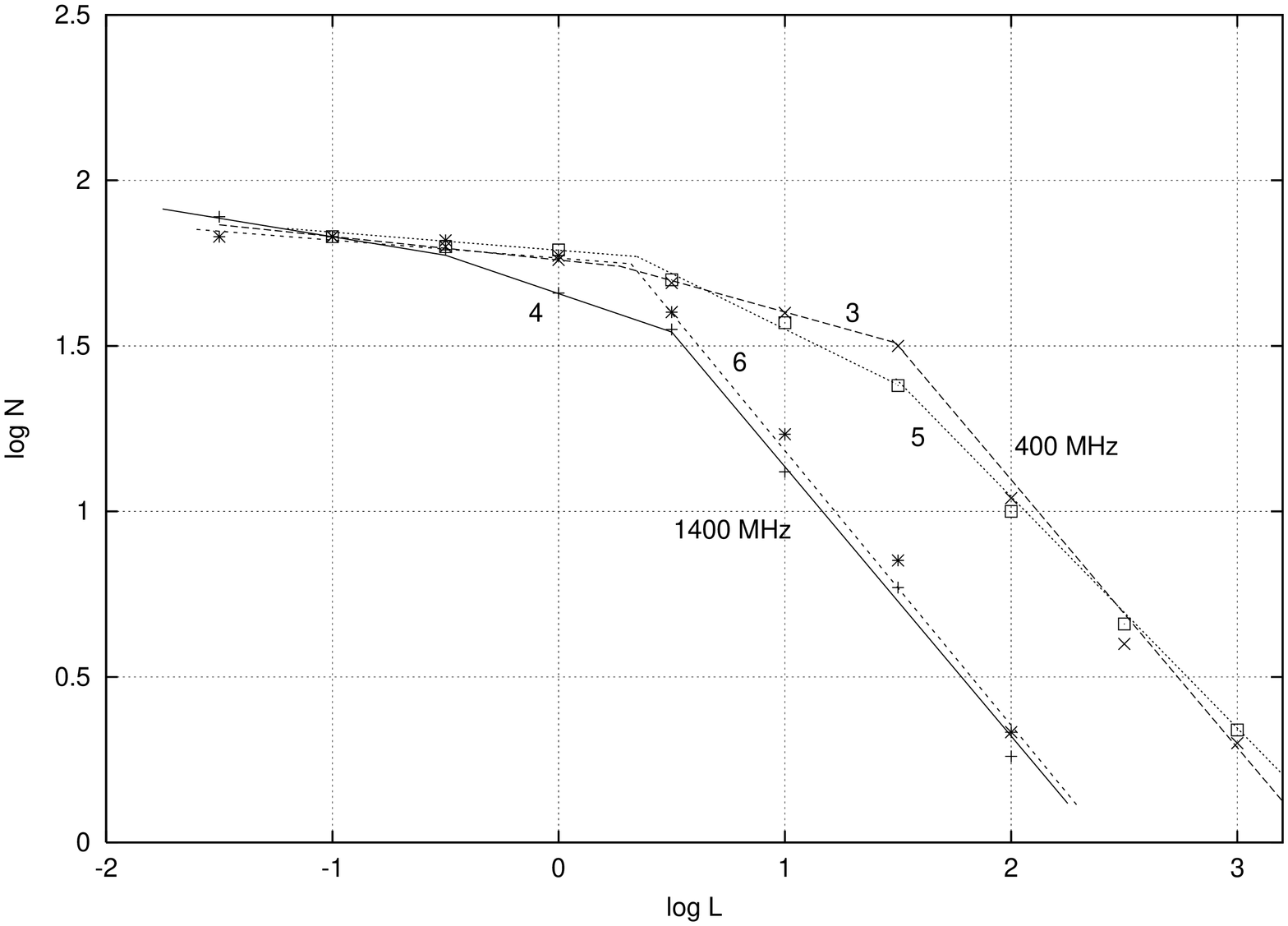,width=13cm,angle=-90}}
{Figure 7. All Luminosity functions together.}
\end{figure*}

\begin{figure*}
\centerline{\psfig{file=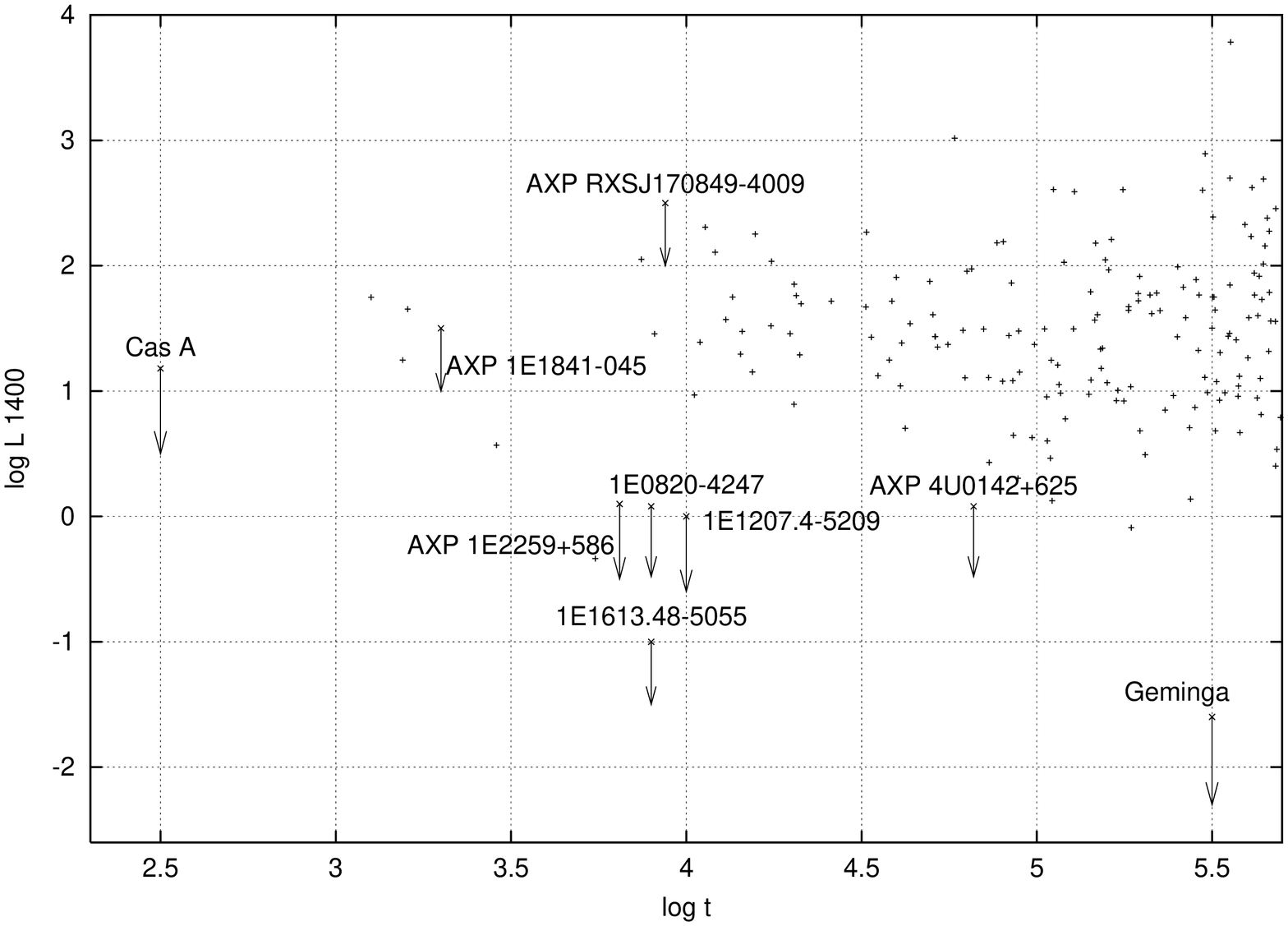,width=13cm,angle=-90}}
{Figure 8. The radio luminosity versus characteristic ages (Log
L$_{1400}$-Log $\tau$) for young PSRs and for different types of single
point X-ray sources.}
\end{figure*}  

\begin{figure*}
\centerline{\psfig{file=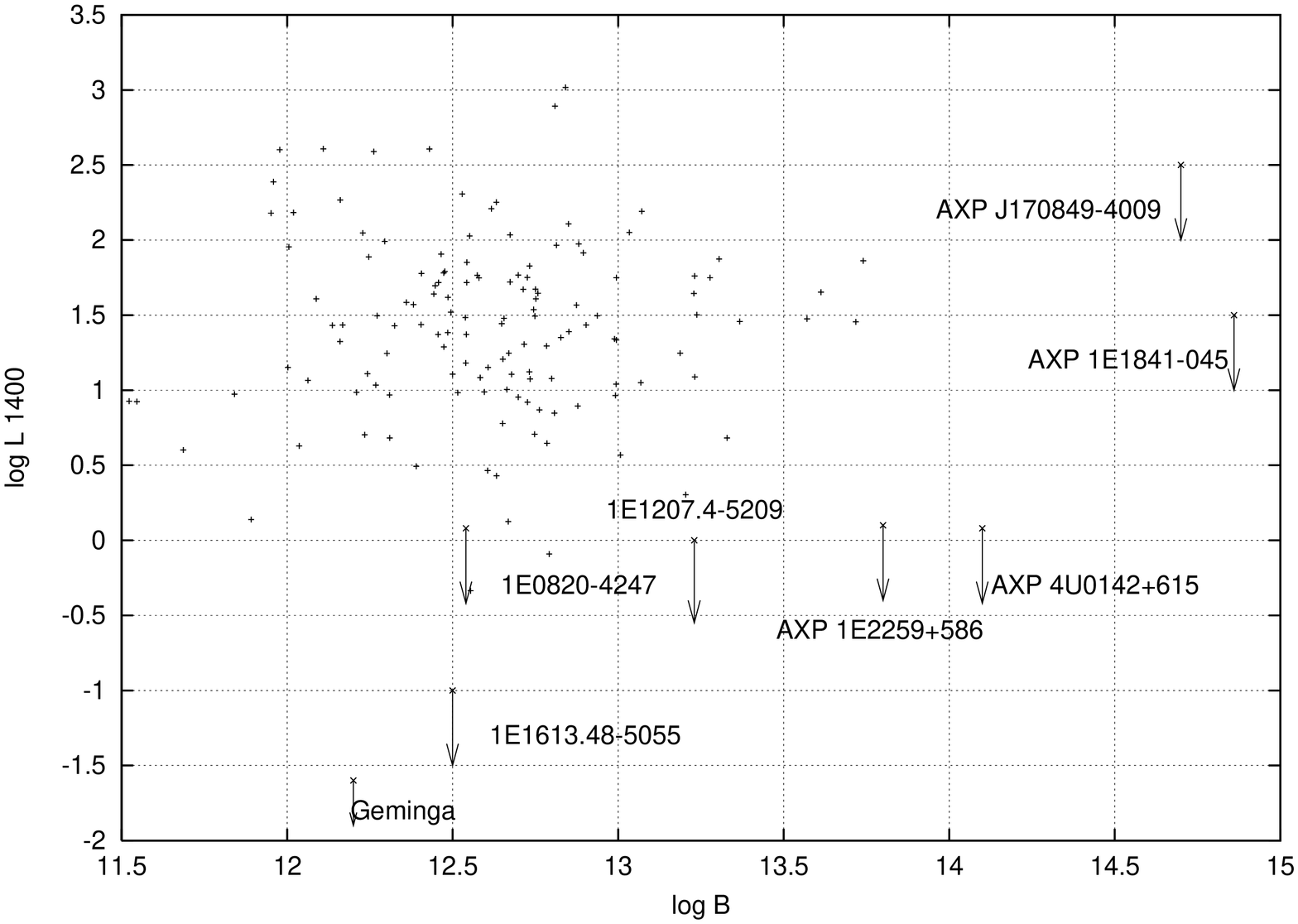,width=13cm,angle=-90}}
{Figure 9. The radio luminosity versus magnetic field strengths (Log
L$_{1400}$-Log B) for young PSRs ($\tau < 3$ 10$^5$ yr) and
different types of single point X-ray sources.}
\end{figure*}

\end{document}